# Firewalls, black-hole thermodynamics, and singular solutions of the Tolman-Oppenheimer-Volkoff equation

W. H. Zurek
Theoretical Astrophysics, California Institute of Technology, Pasadena, California 91125

Don N. Page
Department of Physics, The Pennsylvania State University, University Park, Pennsylvania 16802

We investigate thermodynamic equilibrium of a self-gravitating perfect fluid in a spherically symmetric system containing a black hole of mass $M$ by means of the Tolman-Oppenheimer-Volkoff (TOV) equation. At $r >> 2M$ its solutions describe a Hawking atmosphere with temperature $T_{BH} = (8\pi M)^{-1}$ that is increasingly blueshifted as $r$ approaches $2M$. However, there is no horizon at the Schwarzschild radius. Instead, the fluid becomes increasingly hot and dense there, piling up into a “firewall” with the peak temperatures and densities reaching Planck values somewhat below $r = 2M$. This firewall surrounds a negative point mass residing at $r$=0, the only singularity of the solution. The entropy of the firewall is comparable to the Bekenstein-Hawking entropy.

Solutions of Einstein equations that are spherically symmetric and extremize the entropy of a perfect fluid for fixed total mass satisfy the Tolman-Oppenheimer-Volkoff (TOV) equation[1-4]:

$$\frac{dp(r)}{dr} = \frac{-[\rho(r) + p(r)][m(r) + 4\pi r^3 p(r)]}{r[r - 2m(r)]} \qquad (1)$$

Here $\rho(r)$ and $p(r)$ are the (proper) pressure and density, related to one another by the equation of state of the form $p = f(\rho)$. The *effective mass m(r)* inside a sphere of surface area $4\pi r^2$ is equal to

$$m(r) = 4\pi \int_0^r \rho(r) r^2 dr + m(0) \ . \qquad (2)$$

The TOV equation is a general relativistic version of the well-known equation for hydrostatic equilibrium in a fluid with Newtonian gravity and has been extensively used in the study of relativistic stars[3.4].

The aim of this paper is to use the TOV equation to investigate thermodynamic equilibrium in a spherically symmetric system which contains an uncharged, nonrotating black hole. This equilibrium is possible because the Schwarzschild black hole has the Hawking temperature[5], $T_{BH} = (8\pi M)^{-1}$. Until now this problem has been discussed in the approximation neglecting both the selfgravity of the fluid and the influence of the black-hole metric on its properties[4-10]. Our model is also idealized in that it ignores deviations of the stress-energy tensor from the perfect-fluid form.

We begin our analysis by assuming that the perfect fluid satisfies a $\gamma$-law equation of state

$$p = (\gamma - 1)\rho = \rho/(n-1) \ . \qquad (3a)$$

Its density and specific entropy can be expressed by

$$\rho = \alpha T^{\gamma/(\gamma-1)} = \alpha T^n \ , \qquad (3b)$$

$$s = \alpha\gamma T^{\frac{1}{\gamma}} = [\alpha n/(n-1)]T^{n-1} \ , \qquad (3c)$$

where $T$ stands for the temperature and $\alpha$ is Stefan's constant. To have $0 < p < \rho$, $\gamma$ and $n$ are taken to belong, respectively, to the ranges $1 < \gamma < 2$ and $2 < n < \infty$.

Bondi[11] has noticed that for $\gamma$-law perfect fluids one can simplify Eqs. (1) and (2) by introducing variables

$$u(r) = m(r)/r, \qquad (4a)$$

$$v(r) = 4\pi r^2 p(r) \ . \qquad (4b)$$

In terms of $u(r)$ and $v(r)$ Eqs. (1) and (2) are equivalent to Bondi's equation and an additional equation for the radius:

$$v[2(\gamma - 1) - (5\gamma - 4)u - \gamma u]du$$
$$= (1 - 2u)[v - (\gamma - 1)u]du \ , \qquad (5)$$

$$\frac{dr}{r} = \frac{du}{v/(\gamma - 1) - u} \ . \qquad (6)$$

Bondi’s equation is nonlinear and cannot be solved analytically. For the important case of a fluid consisting of massless quanta, $\gamma = \frac{4}{3}$, numerically obtained solutions $u(v)$ are plotted in Fig. 1. The starlike solution originates at $u = v = 0$. For small $r$ it can be expanded[3] in terms of $y = [4\pi\rho(0)/3]^{1/2} r$
$u(y) = y^2 - 2.4y^4 + \ldots \quad v(y) = y^2 - 4y^4 + \ldots$

In the limiting case $\rho(0) \to \infty$, $m(0) = 0$, the TOV equation gives $p(r) = 1/(56\pi r^2)$, $m(r) = \frac{3}{14}r$. The oscillatory approach to the focus at $u = \frac{3}{14}$, and $v = \frac{1}{14}$ is reflected in the "damped oscillations" of $p(r)$, $\rho(r)$, and $m(r)$ around that limiting solution with the increase of $\rho(0)$. All solutions that are regular at the origin have $u < \frac{1}{14}$, and so $r > 2[2m(r)]$ for all $r$[3,12]. Starlike equilibria of $\gamma = \frac{4}{3}$ fluid have been reexamined in the paper stimulated by Bekenstein's suggestion[13] for the limit on the entropy of a finite system, $S < 2\pi RE$, by Sorkin, Wald, and Jiu[14]. They conclude that the entropy of an everywhere regular solution of the TOV equation is never greater than $R^{3/2}$, which is by a factor $R^{1/2}$ less than that limit. Singular solutions of the TOV equation have not been described before[15].

In the case of the star it was customary and convenient to integrate the TOV equation starting from $r = 0$. There it suffices to give only the central density $\rho(0)$ as the mass $m(0) = 0$. The integration, with some definite equation of state, provides $m(r)$ and $\rho(r)$. This procedure cannot be followed in the case investigate here, as there is a black hole at the center of the system. Therefore, we shall integrate TOV equation inwards, starting at some finite $r = R$.

In the case considered here it is convenient to give initial conditions near the origin of the $u(v)$ graph. The solution of interest is close to the starlike solution, but misses the origin by bending upwards, as shown in Fig. 1a: For a given $v \ll 1$, $u$ is bigger than that of the starlike solution. Note the effective mass decreasing to an approximately constant value $M$---corresponding to a mass of the central black hole---from the initial $m(R)=E$. This decrease is caused by the fluid with approximately constant density. When the radius approaches the estimated location of the horizon, the density of the fluid increase dramatically to $\rho_{\max} \approx 1$, while the effective mass becomes negative. It is surprising and rather unexpected to see the solution continue through $r=2M$ without encountering any horizon. The integration can be extended to the origin. There resides a negative point mass, the only singularity of the above solution.

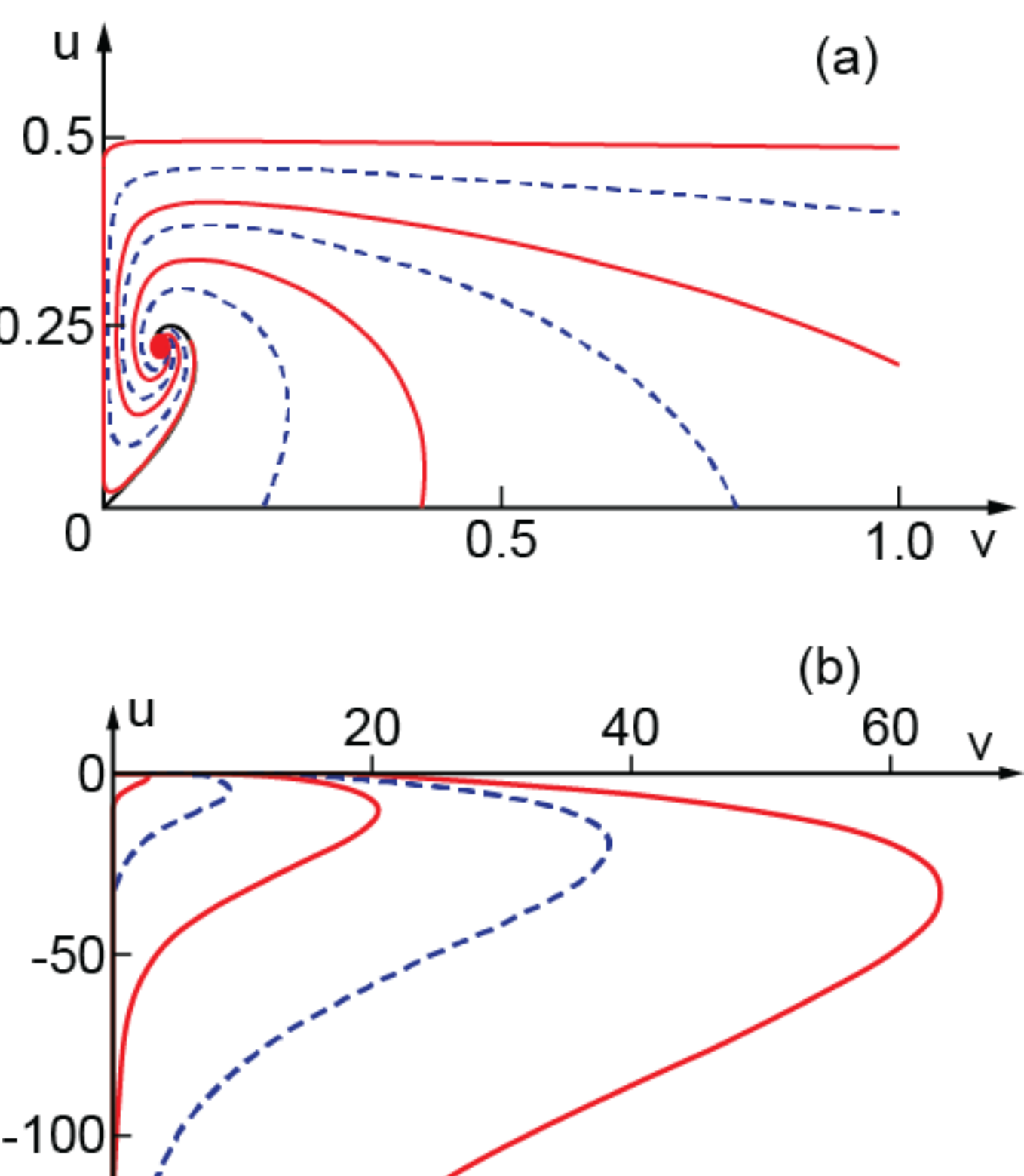

Fig. 1. Solutions of Bondi's equation for a fluid of massless quanta $\gamma = \frac{4}{3}$, $n$=4. Physically interesting solutions satisfy $u(r) = m(r)/r < \frac{1}{2}$; $v(r) = 4\pi r^2 p(r) \geq 0$. (a) For $u > 0$, Bondi's equation exhibits a focus at $v = \frac{1}{14}$ and $u = \frac{3}{14}$. The solution which originates at $u = v = 0$ is starlike. Singular solutions with $M \gg 1$ can be obtained analytically in four overlapping regions: (i) $u \ll 1$, $v \ll 1$, (ii) $v \ll (\gamma - 1)u$, (iii) $1 - 2u \ll 1$, and (iv) $v + |u| \gg 1$. These approximate solutions can be smoothly connected by adjusting integration constants. Part of this last range (iv) is shown in (b). Note the scale change between (a) and (b).

The entropy of the fluid inside $r=2M$ computed for a fluid of massless quanta, equals $\sim 1.6(4\pi M^2)$. This behavior---no causal horizon and a negative point mass at $r = 0$ ---is generic for solutions of the TOV equation with the $\gamma$ -law equation of state that originate in the physically accessible region: $v > 0$, $0 < u < \frac{1}{2}$.

Further analysis of this TOV model for the equilibrium between the black hole and a perfect fluid can be achieved by solving Eqs. (5), (6) analytically. Again we begin in the range $u \ll 1$, $v \ll 1$.

There, Eq. (5) simplifies to

$$2udu \cong [v/(\gamma-1)-u]dv .$$

Approximate solutions are, therefore, $v = 4\pi r^2 P$ and $u = M/r + v/3(\gamma-1)$, where $P$ and $M$ are constants and have the obvious physical interpretation. The temperature is also constant and given by $T = \left[P/\alpha(\gamma-1)\right]^{1-1/\gamma}$. It can be regarded as the Hawking temperature of the black hole of mass $M$, if we set $T = T_{BH} \equiv (8\pi M)^{-1}$ or $P = (\gamma-1)\alpha(8\pi M)^{-\gamma/(\gamma-1)}$. The energy and entropy densities are approximately constant wherever $u \ll 1$, i.e. for all

$$2M \ll r \ll (\gamma-1)^{1/2} P^{-1/2} = \alpha^{-1/2}(8\pi M)^{\gamma/(2\gamma-2)} .$$

As $r$ decreases towards $2M$, $u(r)$ gets near ½ while $v(r)$ initially remains small when

$$(8\pi M)^2 P = (\gamma-1)\alpha(8\pi M)^{-(2-\gamma)/(\gamma-1)} \ll \gamma - 1 .$$

Thus,

$$v \ll (\gamma-1)u$$

and

$$v[2(\gamma-1)-(5\gamma-4)u]du \cong -(1-2u)(\gamma-1)u dv ,$$

$$dr/r \cong -du/u .$$

Functions

$$u(r) = M/r$$

and

$$v(r) \cong 4\pi r^2 P[1-2u(r)]^{-\gamma/(2\gamma-2)}$$

solve these equations and merge with the previous solutions in the overlap region, $v \ll u \ll 1/2$. They represent fluid with blue-shifted temperature $T \cong T_{BH}/[1-2u(r)]^{1/2}$. In the range $2M < r < 8\pi M$ the dominant wavelength of the quanta of black hole radiation becomes comparable to the curvature radius. The fluid can no longer be regarded as perfect. Quantum effects, e.g. vacuum polarization, are likely to play an important role, and the validity of the TOV equation can be questioned. For the time being we shall disregard this and continue our search for the black hole – like solutions of the TOV equation with the $\gamma$-law perfect fluid.

In the range $r \approx 2M$, $u \approx 1/2$. There it is helpful to write Bondi's equations in terms of $h = g_{rr}^{-1} = 1-2u(r)$. For $h \ll 1$, arbitrary $v$, one obtains

$$v(2v+1)dh \cong 2h(2v/\gamma - 1 + 1/\gamma)dv ,$$

$$dr/r \cong -(2-2/\gamma)h\, dv/(v+2v^2) .$$

The solution of Eq. (5)

$$h \cong (16\pi P M^2)^{2-2/\gamma}(1+2v)^2 v^{-2+2/\gamma} ,$$

can be expressed as a relation between the radial and temporal component of the metric tensor: $-g_{rr}g_{00} = (1-2v)^{-2}$. The radius changes very little while $h << 1$.

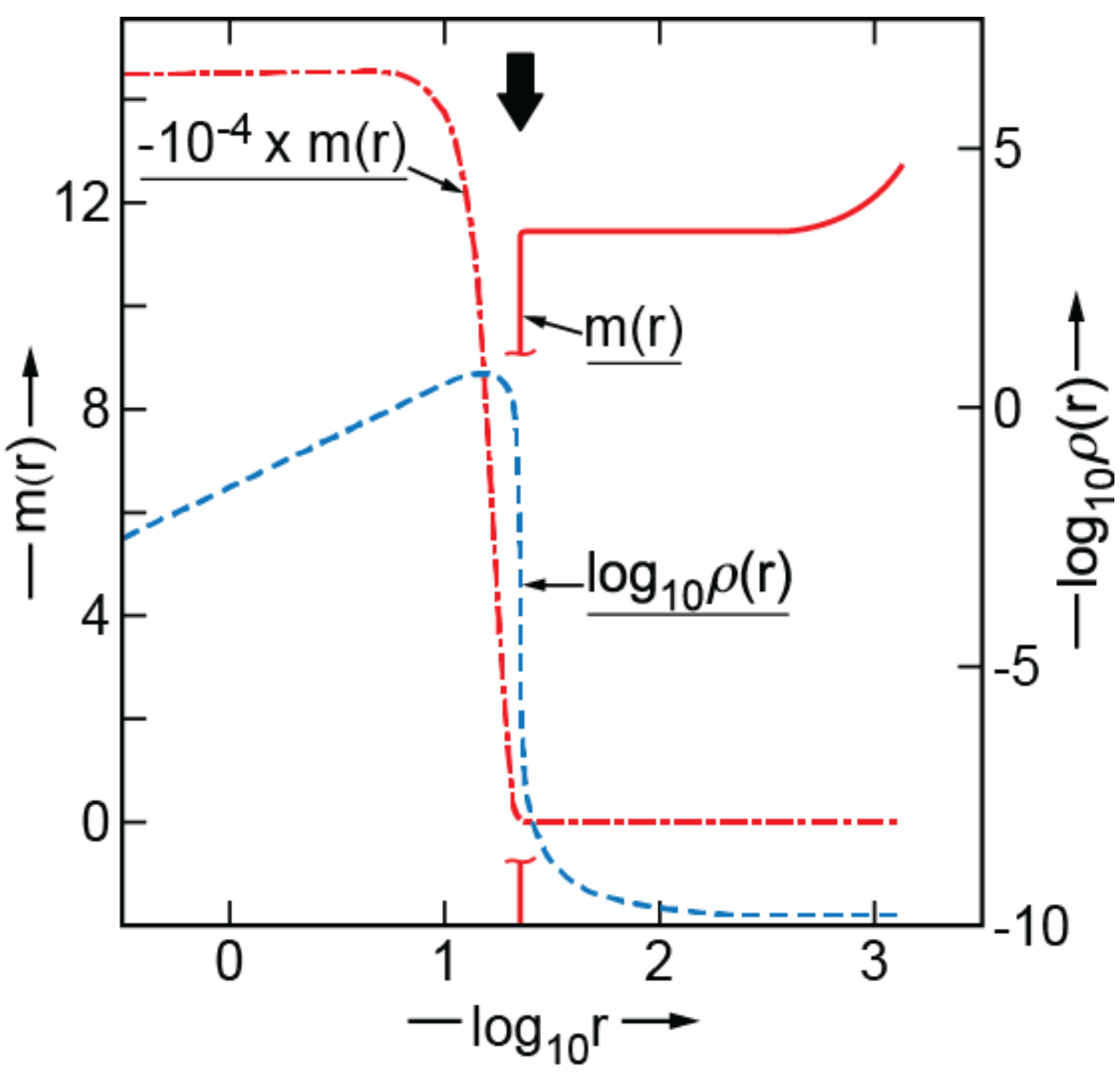

FIG. 2. Effective mass $m(r)$ and density $\rho(r)$ as a function of radius for fluid of massless quanta ($\gamma = 4/3$, $n$=4). Radius, mass, and density are given in absolute (Planck) units. The initial values of $u$ and $v$ were obtained from the condition for thermodynamic equilibrium between a Schwarzschild black hole of mass $M$ and a radiation heat bath confined to a spherical cavity of radius R: $m(r) - M = (4\pi R^3/3)\alpha T_{BH}^4$. The total energy $m(R)$ of the configuration was chosen to be 12.65 Planck units. For simplicity, Stefan's constant $\alpha = 1$. Temperature is assumed to be $T = T_{BH} \equiv (8\pi M)^{-1}$. The heavy arrow points to the expected location of the black hole horizon at $r = 2M$. A negative point mass at $r$=0 is responsible for the negative effective mass inside $r$=$2M$. The density remains positive for all r>0.

As $r$ decreases below $2M$, first $v$ becomes large while $h$ is still small, and then integrating (2) inward through the huge energy density makes $m(r)$ go negative so that $h$ becomes large and stay large down to $r$=0. In this range, $v + h >> 1$, Eq. (5) can be approximated by

$$\frac{dv}{dh} \cong -\frac{v(5\gamma - 4 - 2\gamma v/h)}{h(2\gamma - 2 + 4v/h)} .$$

It can be solved by introducing the auxiliary variable $z = v/h$. For $\gamma < 2$ the solution can be expressed in terms of the variable $x = (r/2M)^{1/(2\gamma-2)}$ and two coefficients: $A = (32\pi PM^2)^{1/(2-\gamma)}$ and $B = \frac{2-\gamma}{7\gamma-6}$. The proper pressure and effective mass are:

$$p(r) \cong PA^{-\gamma}x^{\gamma}(1-x^{7\gamma-6})^{\gamma/(2-\gamma)} ,$$

$$m(r) \cong m(0)(1-x^{7\gamma-6})^{2/(2-\gamma)} + r/2 .$$

The central negative "bare mass" is

$$m(0) \cong -MA^{-2\gamma+2}B^{2/(2-\gamma)}$$

$$\approx -64\pi^2 M^3[(\gamma-1)\alpha/2\pi]^{-2(\gamma-1)/(2-\gamma)}B^{2/(2-\gamma)} .$$

The density reaches maximum at

$$r = 2M[(2-\gamma)/(6\gamma-4)]^{2(\gamma-1)/(7\gamma-6)} .$$

When the pressure $P$ is calculated from Eq. (3) with $T = 1/8\pi M$, the maximal density is of the order of Planck density and depends on $\gamma$, but not on the black hole mass $M$.

The entropy of the interior portion of the above solution is of obvious interest. We calculate it using the formula $S = 4\pi\int_0^{2M} s g_{rr}^{1/2} r^2 dr$. This yields

$$S \cong \left[\frac{2\gamma}{7\gamma-6}\right]\frac{M}{T_{BH}} = \left[\frac{4\gamma}{7\gamma-6}\right]S_{BH} = \left[\frac{4n}{n+6}\right]S_{BH} . \quad (7)$$

For perfect fluid in which the speed of sound approaches the speed of light – in the limit of $\gamma = 2$ or $n$=2 – the entropy inside $r$=$2M$ coincides with the entropy $S_{BH} = 4\pi M^2$ of the black hole, though our approximate solution of the TOV equation must be modified in this limit. Therefore, this fluid configuration attains entropy of the order of the upper limit proposed by Bekenstein[13]. Preliminary analysis indicates that in spite of this large entropy the above configuration may be unstable, and, therefore, a local minimum rather than maximum of entropy[16].

The above "interior solution" is qualitatively different from the Schwarzschild solution for $r$<$2M$. To begin with, the Schwarzschild horizon was not crossed even though $r$<$2M$: Inequality $r$<$2m(r)$ was satisfied for all $r$. Moreover, the central singularity corresponds to a negative, rather than positive, point mass. A shell of very dense fluid $\rho \approx 1$ surrounds this central, negative "bare" mass $m(0)$. The mass of the shell is $M - m(0)$. From the "outside", $r > 2M$, the central object will appear—to an observer stationary with respect to the radiation—to be a Schwarzschild black hole immersed in radiation of the Hawking temperature.

Near $r = 2M$ the TOV equation predicts that the temperature of the fluid will be significantly blue-shifted, in accord with Tolman's formula: $T(r) = T_{BH}/(-g_{00})^{1/2}$. In this near-horizon range it is, however, doubtful whether the usual thermodynamic concepts can be used, as (1) the dominant wavelength of the quanta which constitute the fluid becomes comparable with the scale over which metric varies significantly, (2) $g_{rr}$ and $g_{tt}$ vary by orders of magnitude over distances of the order of the Planck length, and (3) the density of the fluid becomes comparable with the Planck density. In each of the above-mentioned conditions, vacuum polarization and/or quantum gravity can be expected to play a crucial role[17]. In this sense, our results indicate that the origins of black-hole thermodynamics are intimately related to the quantum nature of the fields involved.

We would like to thank John Archibald Wheeler for encouragement at the early stages of this work, and for useful discussions in its course. The help of Bog Kuszta in both numerical and analytic treatment of nonlinear equations was invaluable. Financial support was provided by the NSF Grants Nos. AST-79-22012-Al and PHY81-17464, a Tolman Fellowship (WHZ), and an Alfred P. Sloan Research Fellowship (DNP).